\documentclass[CRPHYS,Unicode,manuscript]{cedram}

\usepackage{amssymb}

\renewcommand{\vec}[1]{\boldsymbol#1}

\title{Quantum transport in strongly correlated Fermi gases}

\alttitle{Transport quantique dans des gaz de fermions fortement corrélés}

\author{\firstname{Tilman} \lastname{Enss}\CDRorcid{0000-0002-5334-2448}}
\address{Institute for Theoretical Physics, University of Heidelberg, Germany}
\thanks{This work is supported by the Deutsche Forschungsgemeinschaft
  (DFG) via Project-ID 273811115 (SFB 1225 ISOQUANT) and under Germany's Excellence Strategy EXC 2181/1-390900948 (the Heidelberg STRUCTURES Excellence Cluster).}
\keywords{Strongly correlated fermions, quantum transport, bulk viscosity,
  hydrodynamics, attractors}

\altkeywords{fermions fortement corrélés, transport quantique,
  viscosité de volume, hydrodynamique, attracteur}

\begin{abstract}
  Transport in strongly correlated fermions cannot be understood by
  fermionic quasiparticles alone.  We present a theoretical framework
  for quantum transport that incorporates strong local correlations of
  fermion pairs.  These contact correlations add essential
  contributions to viscous, thermal and sound transport coefficients.
  The bulk viscosity, in particular, receives its dominant
  contribution from pair excitations.  Moreover, it can be measured
  elegantly by observing the response to a time-dependent scattering
  length even when the fluid is not moving.  Rapid changes of the
  scattering length drive the system far out of local equilibrium, and
  we show how it relaxes back to equilibrium following a hydrodynamic
  attractor before a Navier-Stokes description becomes valid.  This
  paper summarizes a talk given at the Symposium ``Open questions in
  the quantum many-body problem'' at the Institut Henry Poincaré,
  Paris, in July 2024.
\end{abstract}

\begin{altabstract}
Le transport dans les fermions fortement corrélés ne peut être compris
par les seules quasiparticules fermioniques. Nous présentons un cadre
théorique pour le transport quantique qui incorpore les fortes
corrélations locales des paires de fermions. Ces corrélations de
contact ajoutent des contributions essentielles aux coefficients de
transport visqueux, thermique et sonore. La viscosité de volume, en
particulier, reçoit sa contribution dominante des excitations de
paires. En outre, elle peut être mesurée de manière élégante en
observant la réponse à une longueur de diffusion dépendant du temps,
même lorsque le fluide n'est pas en mouvement. Des changements rapides
de la longueur de diffusion éloignent le système de l'équilibre local,
et nous montrons comment il retourne à l'équilibre en suivant un
attracteur hydrodynamique avant qu'une description de Navier-Stokes ne
devienne correcte. Cet article résume un exposé donné au symposium
«Open questions in the quantum many-body problem» à l'Institut Henri
Poincaré, Paris, en juillet 2024.
\end{altabstract}

\begin{document}

\maketitle

\section{Introduction}

Resonantly interacting fermions are characterized by strong
short-range correlations between $\uparrow$ and $\downarrow$ fermions
(see Yvan Castin's presentation in this volume). These correlations
give rise to remarkable transport properties that have been observed
in experiments with ultracold Fermi gases in recent years. Noteworthy
examples include (i) dilute clouds of opposite spin bounce off one
another and create shock waves before they eventually merge
diffusively \cite{sommer2011a}; (ii) the unitary Fermi gas exhibits
extremely low friction, given by the ratio of shear viscosity to
entropy density $\eta/s \gtrsim 0.5\,\hbar/k_B$, and thereby
constitutes a nearly perfect fluid \cite{schaefer2009, cao2011,
  enss2011}; (iii) a quantum lower bound on diffusivity
$D\gtrsim \hbar/m$ is observed for spin diffusion (i)
\cite{sommer2011a, enss2012spin, bardon2014, trotzky2015, luciuk2017,
  enss2019spin} and momentum diffusion (ii) but also for thermal and
sound diffusion \cite{braby2010, patel2020, bohlen2020, frank2020,
  li2022, wang2022, yan2024, li2024universal}; (iv) several transport
relaxation rates $\tau^{-1} \sim k_BT/\hbar$ scale proportional to
temperature in the normal state above the superfluid critical
temperature $T_c$, reminiscent of quantum critical scaling
\cite{nikolic2007, enss2012crit, enss2019spin, frank2020}.

Important questions include how this collective behavior arises from
the microscopic Hamiltonian and how to derive an effective description
at large scales.  Near equilibrium, hydrodynamics works well as an
effective description in the strongly correlated regime that is
dominated by frequent collisions.  However, dissipative hydrodynamics
requires the equation of state and the transport cofficients as input,
and their computation from first principles remains a challenging
task.  Explicit computations have shown quantum limited diffusion in
many instances, but a universal many-body mechanism for different
microscopic models has not yet emerged.  Beyond hydrodynamics, the
short-time behavior and the approach to equilibrium can exhibit
relaxation phenomena on different scales, for instance attractor
behavior beyond a Navier-Stokes description \cite{fujii2024,
  mazeliauskas2025}.

\section{Boltzmann kinetic theory}

The dilute two-component Fermi gas is described by the Hamiltonian
\cite{zwerger2012}
\begin{align}
  \hat H = \int d^dx\, \sum_{\sigma=\uparrow,\downarrow}
  \psi_\sigma^\dagger(\vec x) \left( -\frac{\hbar^2\nabla^2}{2m} -
  \mu_\sigma \right) \psi_\sigma^{\phantom\dagger}(\vec x)
  + g_0 \int d^dx\, \psi_\uparrow^\dagger(\vec x)
  \psi_\downarrow^\dagger(\vec x)
  \psi_\downarrow^{\phantom\dagger}(\vec x)
  \psi_\uparrow^{\phantom\dagger}(\vec x) 
\end{align}
for nonrelativistic fermions of mass $m$ with an attractive
short-range (contact) interaction. The bare coupling strength
$g_0 = [(4\pi\hbar^2a/m)^{-1} - m\Lambda/(2\pi^2\hbar^2)]^{-1}$ in
three dimensions is given in terms of the low-energy $s$-wave
scattering length $a$ and a large-wavenumber cutoff $\Lambda$.  In the
following we set $\hbar=1$.

The first approach to transport in a Fermi gas is by Boltzmann kinetic
theory \cite{smith1989}. The single-particle distribution function
$f(\vec r,\vec p,t)$ evolves according to the Boltzmann equation
\begin{align}
  \frac{\partial f}{\partial t} + \vec v_{\vec p}\cdot \nabla_{\vec r}
  f + \vec F\cdot \nabla_{\vec p} f = \left(\frac{\partial f}{\partial
  t}\right)_\text{coll},
\end{align}
where the left-hand side is the streaming term that includes mean-field
interactions, while the right-hand side denotes the collision integral
\begin{align}
  \label{eq:coll}
  \left(\frac{\partial f_1}{\partial t}\right)_\text{coll}
  \simeq -\int d\vec p_2\, d\Omega\, \frac{d\sigma}{d\Omega}
  \lvert \vec v_1-\vec v_2\rvert 
  \left[ f_1 f_2 (1-f_{1'}) (1-f_{2'}) - (1-f_1) (1-f_2) f_{1'} f_{2'}
  \right].
\end{align}
The collision integral describes how scattering between two particles
$1$, $2$ into new states $1'$, $2'$ leads to a loss (first term) or
gain (second term) of particles in state $1$. At high temperatures
above the Fermi temperature ($T\gg T_F$) the resonant cross section
$d\sigma/d\Omega = 4\hbar^2/\lvert \vec p_1-\vec p_2\rvert^2$ is so
simple that the collision integral can be computed analytically. In
the degenerate Fermi gas ($T\lesssim T_F$) Pauli blocking of final
states reduces the Fermi distribution factors in the collision
integral.  At the same time, however, Pauli blocking of the
intermediate virtual states between scatterings enhances the cross
section $d\sigma/d\Omega$ \cite{bruun2005, enss2012crit}.  Near the
scattering resonance, remarkably these two competing effects cancel
almost perfectly and the resulting collision rate $\tau^{-1}$ follows
nearly classical scaling \cite{bruun2009, frank2020}.  The relaxation
time $\tau$ is then combined with thermodynamics (in the case of shear
viscosity, the pressure $p$) to yield the frequency dependent
transport coefficient, for instance the complex shear viscosity
$\eta(\omega) = p\tau_\eta/(1-i\omega\tau_\eta)$ as follows from the
memory function formalism \cite{frank2020, mandal2024}. The Boltzmann
prediction for sound attenuation is found to agree with experimental
data \cite{patel2020} for the degenerate unitary gas down to
$T\simeq 2T_c$, which constitutes a remarkable success of kinetic
theory in the strongly correlated regime. While finite-range
corrections to transport are subleading for $s$-wave interactions,
they can give rise to prominent effects such as quasi-bound states for
$p$-wave interactions \cite{maki2023}.

In the collision integral \eqref{eq:coll} the two-particle
distribution function has been factorized into the product of two
separate distribution functions for particles $1$ and $2$.  This
factorization is based on the assumption of molecular chaos and does
not capture the strong local pair correlations
$g_{\uparrow\downarrow}^{(2)}(r) \sim \mathcal C/r^2 + \mathcal
O(1/r)$ at short distance, where $\mathcal C$ denotes the expectation
value of the contact operator (see below).  In the following we will
see how these short-range correlations affect transport.

\section{Kubo formula and bulk viscosity}

A more general approach to transport, which makes no quasiparticle
assumption, is derived in linear response theory. The transport
coefficients are related by Kubo formulas to equilibrium expectation
values; for instance the frequency dependent shear viscosity is given
in terms of the transverse stress response function \cite{taylor2010,
  enss2011},
\begin{align}
  \label{eq:eta}
  \eta(\omega) = \int d^dx\, dt\,
  \frac{e^{i(\omega+i0)t}-1}{i(\omega+i0)}\,
  i\theta(t) \langle [\hat\Pi_{xy}(\vec x,t), \hat\Pi_{xy}(0,0)] \rangle .
\end{align}
The microscopic expression for the shear stress operator is
$\hat\Pi_{xy}(\vec x) = \frac1{2m} \sum_\sigma \bigl[\partial_x
\hat\psi_\sigma^\dagger(\vec x) \partial_y \hat\psi_\sigma(\vec x) +
\partial_y \hat\psi_\sigma^\dagger(\vec x) \partial_x
\hat\psi_\sigma(\vec x)\bigr] - \int d^3r\,
\frac{r_xr_y}r\,\frac{\partial v(r)}{\partial r}
\hat\psi_\uparrow^\dagger(\vec x+\vec r/2)
\hat\psi_\downarrow^\dagger(\vec x-\vec r/2) \hat\psi_\downarrow(\vec
x-\vec r/2) \hat\psi_\uparrow(\vec x+\vec r/2)$ for short-range
potential $v(r)$ \cite{martin1959}. It has two contributions: the
first, quadratic term gives the main contribution for gases, while the
second, quartic term dominates in fluids \cite{enss2011}. Furthermore,
the bulk viscosity $\zeta$ characterizes friction during isotropic
expansion and contributes to sound attenuation. In constrast to the
shear viscosity, however, the bulk viscosity is constrained by
symmetry and vanishes identically for a scale invariant gas such as
the ideal gas but also for the unitary Fermi gas \cite{werner2006,
  son2007}. It can be computed by the Kubo formula \cite{fujii2020}
\begin{align}
  \label{eq:zeta}
  \zeta(\omega) = \int d^dx\, dt\,
  \frac{e^{i(\omega+i0)t}-1}{i(\omega+i0)}\,
  i\theta(t) \langle [\delta \hat p(\vec x,t), \delta \hat p(0,0)] \rangle 
\end{align}
in terms of the operator $\delta \hat p$ that measures pressure
fluctuations.  The pressure $p=-\partial E/\partial V$ is obtained by
performing a scale transformation, and specifically for the dilute gas
in three dimensions one obtains the pressure operator
\begin{align}
  \hat p = \frac23 \hat{\mathcal H} + \frac{\hat{\mathcal C}}{12\pi m a},
\end{align}
where $\hat{\mathcal H}$ denotes the Hamiltonian density and $a$ the
scattering length.  The last term involves on the contact operator
\begin{align}
  \hat{\mathcal C}
  = m^2g_0^2\hat n_\uparrow(\vec x) \hat n_\downarrow(\vec x)
  = \hat\Delta^\dagger(\vec x) \hat\Delta(\vec x),
\end{align}
which is the continuum version of the doublon or pair density
regularized by the bare coupling $g_0 \sim -r_0$ such that its
zero-range limit $r_0\to0$ is well defined
\cite{werner2012}. Equivalently, the contact operator can be expressed
in terms of the local pair operator
$\hat\Delta = mg_0\hat\psi_\downarrow\hat\psi_\uparrow$, such that the
contact measures the density of local pairs.  At the scattering
resonance $1/a=0$ the scale invariant pressure relation
$p=(2/3)\mathcal E$ is recovered, while the contact term quantifies
the deviation from scale invariance due to pairing fluctuations. The
pressure fluctuations are now given as the component of the pressure
orthogonal to density and energy fluctuations \cite{fujii2020},
\begin{align}
  \delta \hat p = \hat p - (\partial p/\partial n)_{\mathcal E} \hat n
  - (\partial p/\partial \mathcal E)_n \hat{\mathcal H}.
\end{align}
Because conserved quantities do not contribute to dissipation, in the
dynamical response the dissipation at $\omega>0$ can only arise from
the response function of the contact operator,
$\delta \hat p = \hat{\mathcal C}/(12\pi m a)$, which is not
conserved. One thus finds that the bulk viscosity at nonzero frequency
is given by \cite{nishida2019, enss2019bulk, hofmann2020, fujii2020}
\begin{align}
  \zeta(\omega>0) = \frac{1}{(12\pi ma)^2} \int d^dx\, dt\,
  \frac{e^{i(\omega+i0)t}-1}{i(\omega+i0)}\, i\theta(t)
  \langle [\hat{\mathcal C}(\vec x,t), \hat{\mathcal C}(0,0)] \rangle .
\end{align}
Hence, bulk viscosity is a pure interaction effect that arises from
fluctuations of the pair density, not of single fermions.  These
contributions are not easy to capture in a fermionic kinetic theory,
even if the interaction functional is included \cite{dusling2013}.
Instead, the bulk viscosity can be computed using self-consistent
conserving approaches (Luttinger-Ward), which are formulated in terms
of coupled fermion and pair degrees of freedom \cite{haussmann2007,
  enss2011, enss2012spin, enss2019bulk}.  Explicit microscopic
computations for the contact correlations and bulk viscosity in the
degenerate, strongly correlated gas \cite{enss2019bulk} show a
low-frequency Drude peak in the complex bulk viscosity
$\zeta(\omega) \simeq \chi\tau_\zeta/(1-i\omega\tau_\zeta)$
followed by an anomalous contact tail
$\zeta(\omega\to\infty) \sim C/\omega^{3/2}$ at large frequency.
Remarkably, in the unitary gas the bulk scattering rate
$\tau_\zeta^{-1} \propto T$ exhibits a $T$-linear scaling in a wide
temperature range from slightly above $T_c$ to high temperatures above
$T_F$, in distinction to other transport relaxation rates that decay
at high temperatures. This unusual scaling arises from scattering
between pairs, not individual fermions, and is specific to the bulk
viscosity.

Open questions concern the response in the low-temperature, superfluid
state, where a superfluid of fermion pairs can behave differently from
a bosonic superfluid due to the additional pair-breaking excitations
\cite{einzel1984, kurkjian2019}. In particular for the bulk viscosity,
but also for the other transport coefficients it is desirable to
derive a kinetic theory that captures the strong fermion
correlations. A kinetic formulation in terms of coupled fermions and
pairs has been derived in the high-temperature virial expansion
\cite{fujii2023}: the fermionic contribution to the total bulk
viscosity agrees with previous Boltzmann calculations
\cite{dusling2013}, but the pair contribution is found to be much
larger near unitarity. Efforts are underway to extend this to the
quantum degenerate regime.

\subsection{Measurement of the bulk viscosity}

Often transport measurements observe the damping of fluid motion:
elliptic flow or quadrupole motion for shear viscosity, and isotropic
flow or radial breathing for the bulk viscosity.  The measurement of
sound attenuation \cite{patel2020, bohlen2020, li2022, wang2022,
  yan2024, li2024universal, huang2024} gives access to a combination
of several transport coefficients, as sound decays by both momentum
and thermal relaxation processes.  For the bulk viscosity, however,
there is another way of measurement in a dilute gas that works even if
the fluid is homogeneous and at rest \cite{fujii2018}.  In linear
response the contact correlation, and thereby the bulk viscosity, is
given by the response of the contact to an earlier change of
scattering length \cite{enss2019bulk},
\begin{align}
  \label{eq:contactresponse}
  i\theta(t-t') \langle [\hat{\mathcal C}(\vec x,t), \hat{\mathcal C}(0,t')] \rangle
  = -4\pi m\left.\frac{\partial\langle \hat{\mathcal C}(\vec x,t)\rangle}
  {\partial a^{-1}(0,t')}\right\lvert_{S,N}
\end{align}
at fixed entropy and particle number.  Experimentally the spatially
integrated contact has been measured with a high temporal resolution
\cite{bardon2014, luciuk2017}, and one can modulate the scattering
length in time via the applied magnetic field to measure the
response.  In this way, the frequency dependence of the bulk
viscosity can be mapped out.

\section{Attractors to hydrodynamics}

When a system is brought far from equilibrium, one might expect that
the approach to equilibrium at long times is governed by hydrodynamics
(Navier-Stokes equation). In heavy-ion collisions, however, fluid
behavior is found already at short times after a collision, earlier
than hydrodynamics is expected to be valid, in a so-called
hydrodynamic attractor \cite{heller2015}. In general, one can ask
which equations describe the approach to hydrodynamics and whether
they are universal. Furthermore, if hydrodynamics is viewed as a
``derivative expansion'' in powers of $\omega\tau$, what determines
the higher orders? Some answers may be provided by experiments with
ultracold quantum gases, where the time-resolved evolution toward
equilibrium can be observed starting from defined initial conditions
or subject to a particular driving.

\begin{figure}[tbp]
\includegraphics[width=0.66\linewidth]{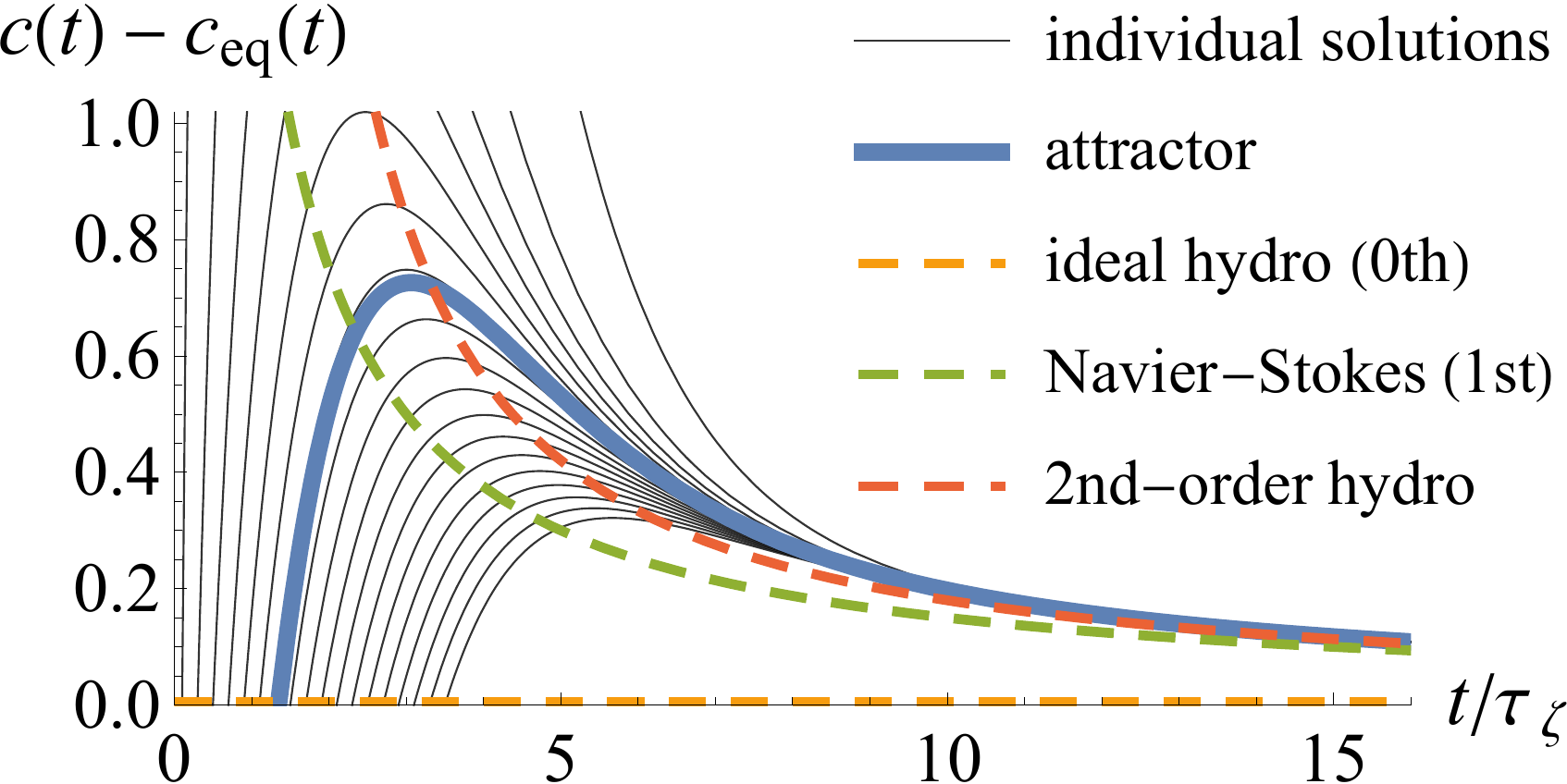}
\caption{Hydrodynamic attractor. The normalized bulk pressure
  $c(t)-c_\text{eq}(t) = \pi(t)/\chi$ exhibits different time
  evolutions for different initial conditions (thin lines), which
  quickly converge toward the attractor solution (thick blue line) and
  only later approach Navier-Stokes hydrodynamics (green dashed
  line). Adapted from \cite{fujii2024}.}
\label{fig:attractor}
\end{figure}

Hydrodynamic attractors can arise in many forms of fluid motion, but
there is a particularly simple case where it can be studied in a
uniform ultracold atomic gas at rest, with no moving parts, when the
scattering length $a(t)$ is ramped at time $t>0$ to bring the system
out of local equilibrium \cite{fujii2018, fujii2024}. The relaxation
back to equilibrium can be observed in the equation of state, most
directly in the contact density expectation value, which is given in
linear response as
\begin{align}
  \label{eq:Ct}
  \mathcal C(t) = \mathcal C_\text{eq} + \int_0^\infty dt'\,
  \frac{\partial \mathcal C(t)}{\partial a^{-1}(t')} \delta a^{-1}(t').
\end{align}
Using Eq.~\eqref{eq:contactresponse} this can be expressed in terms of
the contact correlation, and one finds that the approach to
equilibrium occurs via local dissipation, with the dissipation rate
set by the bulk viscosity \cite{fujii2018}. The Drude peak of the bulk
viscosity \cite{enss2019bulk}
$\zeta(\omega) \simeq \chi\tau_\zeta/(1-i\omega\tau_\zeta)$ (see
above) corresponds in the time domain to an exponential decay of the
contact response within the bulk relaxation time $\tau_\zeta$:
\begin{align}
  \label{eq:Cresponse}
  \frac{\partial \mathcal C(t)}{\partial a^{-1}(t')}
  \simeq \theta(t-t') \left(\frac{\partial \mathcal C}{\partial
  a^{-1}}\right)_{S,N} 
  \frac{\exp[-(t-t')/\tau_\zeta]}{\tau_\zeta}.
\end{align}
By inserting this form into Eq.~\eqref{eq:Ct} one can predict the time
evolution of the contact following arbitrary drives $\delta a^{-1}(t)$
as long as the drive amplitude is small enough to remain in the linear
response regime. But does this time evolution agree with the
prediction of Navier-Stokes hydrodynamics? When the scattering length
is varied, the local pressure also changes in time, and the
nonequilibrium component of the pressure is quantified by the
dissipative bulk pressure
\begin{align}
  \pi(t) = \frac{\mathcal C(t) - \mathcal C_\text{eq}[a(t)]}{12\pi ma(t)},
\end{align}
which for a dilute gas is given in terms of the difference of the
instantaneous contact and the equilibrium contact for the
instantaneous scattering length \cite{fujii2018}.  In Navier-Stokes
hydrodynamics the bulk pressure $\pi = -\zeta V_a$ is driven by the
local expansion of the fluid, $V_a=\nabla\cdot\vec v$, times the bulk
viscosity $\zeta$.  On the other hand, when the scattering length is
changed, the local scale variation arises equally from the rate of
change of the scattering length, $V_a(t) = - 3\dot a(t)/a(t)$.
Therefore, both expansion and variations of the scattering length are
equivalent ways to probe local bulk dissipation.  In contrast to
Navier-Stokes hydrodynamics, we obtain the equation of motion for
$\pi(t)$ from the time derivative of Eq.~\eqref{eq:Cresponse}, which
we have derived microscopically \cite{fujii2024}:
\begin{align}
  \tau \dot\pi(t) + \pi(t) = -\zeta[a(t)] V_a(t). 
\end{align}
This differs from Navier-Stokes by the relaxation term on the
left-hand side, which has the same form as in a
M\"uller-Israel-Stewart formulation.  For a given
external drive $a(t)$ the bulk pressure is obtained by integrating
this differential equation, and the result can be compared to the
Navier-Stokes prediction $\pi_\text{NS}(t)=-\zeta[a(t)] V_a(t)$. For
slow drive frequencies $\omega\tau_\zeta\ll1$ the dissipative term
$\tau\dot\pi$ has little effect and the bulk pressure follows the
drive almost instantaneously. For fast drives
$\omega\tau_\zeta\gtrsim1$, instead, $\pi(t)$ follows the drive with a
time delay and a deviation from Navier-Stokes hydrodynamics is
predicted. This is exemplified by a power-law drive
$a^{-1}(t>t_\text{ini}) \equiv a_\text{ini}^{-1}
(t/t_\text{ini})^{-\alpha}$, which starts at a finite scattering
length and sweeps at first fast, then slower toward unitarity
$a^{-1}=0$. In this case the bulk pressure is found analytically as
\cite{fujii2024}
\begin{align}
  \pi(t) = \pi_\text{ini}\, e^{-(t-t_\text{ini})/\tau_\zeta} +
  \pi_\text{att}(t), \quad
  \pi_\text{att}(t) = c_\alpha\chi\,
  e^{-t/\tau_\zeta} \Gamma(-2\alpha, -t/\tau_\zeta)
\end{align}
in terms of the sum rule $\chi=\zeta/\tau_\zeta$ and the incomplete
Gamma function $\Gamma(s,z)$. The first term describes the exponential
decay of initial conditions on time scale $\tau_\zeta$, while the
so-called hydrodynamic attractor solution $\pi_\text{att}(t)$ is the
same for different initial conditions and depends only on the
transport properties $\zeta$, $\tau_\zeta$ as well as the drive
parameter $\alpha$. For different initial conditions the bulk pressure
is found to first converge toward the attractor solution
$\pi_\text{att}$ before the attractor itself approaches the
Navier-Stokes prediction at longer times, cf.\
Fig~\ref{fig:attractor}.

Standard hydrodynamics is recovered in the solution $\pi(t)$ in the
long-time limit. When expanding in ``temporal gradients''
$\tau_\zeta/t\ll1$, the leading order reproduces $\pi_\text{NS}(t)$,
but the subsequent orders have factorially growing coefficients
$a_n \sim (n+2\alpha)!$ and form an asymptotic series. The initial
condition, furthermore, is nonperturbative in $\tau_\zeta/t$ and is
therefore a nonhydrodynamic mode. Even though the gradient expansion
does not converge, the solution obtained from the equation of motion
is physical and accessible with current experiments
\cite{fujii2024}.

\section{Conclusion}

The short-time attractor behavior in a driven system is an example of
a microscopically motivated extension of hydrodynamics beyond
Navier-Stokes. Cold atom experiments can observe these attractors in
real time by measuring the response of the contact to variations in
the scattering length, thus probing isotropic expansion and local
dissipation by an external drive with no moving parts. A remarkable
prediction of quantum transport theory is that the bulk relaxation
rate $\tau_\zeta^{-1} \propto T$ scales approximately linearly in
temperature but is largely independent of density \cite{enss2019bulk};
this is because pressure fluctuations couple predominantly to pairs
rather than individual fermions. The computation of frequency
dependent transport coefficients remains a challenge, also in the
superfluid state. Recently, the accurate computation of fermion and
pair spectra was achieved by solving the self-consistent
Luttinger-Ward equations directly in real frequency
\cite{johansen2024, enss2024, dizer2024}, which match recent
experiments \cite{li2024}. It will be interesting to extend these
methods and compute dynamical response functions in real frequency,
such as Eqs.~\eqref{eq:eta} and \eqref{eq:zeta}, which determine the
transport coefficients near equilibrium. Interesting questions arise
also in the far-from-equilibrium response, which is strongly affected
by conformal symmetry \cite{maki2022dynamics} and which can
generically be computed using the Keldysh formulation
\cite{bonitz2016}.

\bibliographystyle{crunsrt}


\bibliography{all}

\end{document}